\documentclass[aps,prl,twocolumn,showpacs,superscriptaddress,groupedaddress]{revtex4}

\usepackage{graphicx}  
\usepackage{dcolumn}   
\usepackage{bm}        
\usepackage{amssymb}   
\usepackage{amsmath}
\usepackage{units}
\usepackage{times}

\begin{document}

\title{Origin of Rashba-splitting in the quantized subbands at $\mathbf{Bi}_2\mathbf{Se}_3$ surface}

\author{H.M. Benia}
\email[Corresponding author; electronic address:\ ]{h.benia@fkf.mpg.de}
\affiliation{Max-Planck-Institut f\"ur Festk\"orperforschung, 70569 Stuttgart, Germany}
\author{A. Yaresko}
\affiliation{Max-Planck-Institut f\"ur Festk\"orperforschung,
70569 Stuttgart, Germany}\affiliation{Donostia International Physics Center (DIPC), 20018 San Sebastian/Donostia,
Spain}
\author{A.P. Schnyder}
\affiliation{Max-Planck-Institut f\"ur Festk\"orperforschung,
70569 Stuttgart, Germany}
\author{J. Henk}
\affiliation{Institut f\"ur Physik $-$ Theoretische Physik, Martin-Luther-Universit\"at Halle-Wittenberg, D-06099 Halle (Saale), Germany}
\author{C.T. Lin}
\affiliation{Max-Planck-Institut f\"ur Festk\"orperforschung,
70569 Stuttgart, Germany}
\author{K. Kern}
\affiliation{Max-Planck-Institut f\"ur Festk\"orperforschung,
70569 Stuttgart, Germany} \affiliation{Institut de Physique de la Mati{\`e}re Condens{\'e}e, Ecole Polytechnique
F{\'e}d{\'e}rale de Lausanne, 1015 Lausanne, Switzerland}
\author{C.R. Ast}
\affiliation{Max-Planck-Institut f\"ur Festk\"orperforschung,
70569 Stuttgart, Germany}

\date{\today}

\begin{abstract}
We study the band structure of the $\text{Bi}_2\text{Se}_3$ topological insulator (111) surface using angle-resolved photoemission spectroscopy. We examine the situation where two sets of quantized subbands exhibiting different Rashba spin-splitting are created via bending of the conduction (CB) and the valence (VB) bands at the surface. While the CB subbands are strongly Rashba spin-split, the VB subbands do not exhibit clear spin-splitting. We find that CB and VB experience similar band bending magnitudes, which means, a spin-splitting discrepancy due to different surface potential gradients can be excluded. On the other hand, by comparing the experimental band structure to first principles LMTO band structure calculations, we find that the strongly spin-orbit coupled Bi 6$p$ orbitals dominate the orbital character of CB, whereas their admixture to VB is rather small. The spin-splitting discrepancy is, therefore, traced back to the difference in spin-orbit coupling between CB and VB in the respective subbands' regions.
\end{abstract}

\pacs{79.60.-i, 73.20.-At, 73,21-Fg, 75.70.-Tj}
\maketitle

\begin{figure*}
\centerline{ \includegraphics[width = 0.98\textwidth]{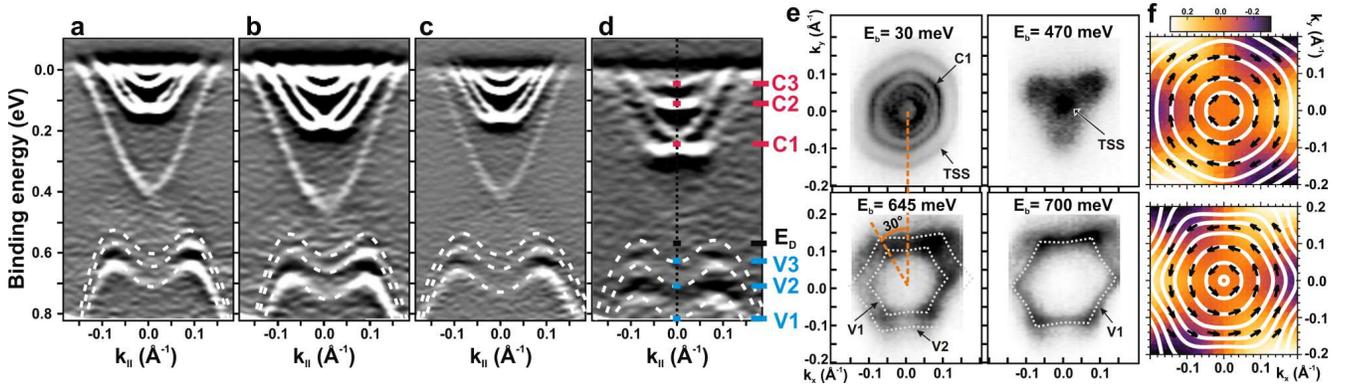}}
\caption{Second derivative of experimental surface band structures from $\text{Bi}_2\text{Se}_3$ crystal at different conditions: (i) Air cleaved crystal after \textbf{a}. two hours at 100 K, \textbf{b}. eight hours at 100 K, and (ii) Vacuum cleaved crystal exposed to \textbf{c}. 720 L and \textbf{d}. 1140 L of water vapor at 100 K (L: Langmuir). \textbf{e}. constant energy cuts at different energies of the band structure in b. The hexagonal warping of the TSS and the outer contour of C1 is turned by $30^{\circ}$ as compared to the warping of the VSBs. All the measurements were performed at 100 K. White dashed lines are guides to the eye. $E_D$, and C1, C2, and C3, and V1, V2, and V3 denote respectively Dirac point, CSBs, and VSBs positions at $k_\| = 0$ $\text{\AA}^{-1}$. \textbf{f}. Calculated energy contours (white bold lines) and in-plane (blue) and out-of-plane (color scale) spin polarization of the TSS above (upper panel) and below (lower panel) $E_D$.}\label{fig:BiSeWaterExp}
\end{figure*}

The narrow bandgap semiconductor $\text{Bi}_2\text{Se}_3$ has been known for decades for its good thermoelectric properties \cite{mishra_electronic_1997,ivanova_extruded_2009}. The recent observation of a topologically protected surface state (TSS) in $\text{Bi}_2\text{Se}_3$ \cite{xia_observation_2009,hsieh_tunable_2009} marked the discovery of a model system for 3D topological insulators (TIs) and has lead to a surge of renewed interest in the properties of this material. The $\text{Bi}_2\text{Se}_3$ (111) surface hosts within a gap of bulk bands projected onto the surface Brillouin zone (BZ) a single TSS with Dirac cone dispersion \cite{xia_observation_2009,hsieh_tunable_2009,zhang_topological_2009-1}. Similar TSS dispersion has been found in other Bi-based systems, such as, $\text{Bi}_2\text{Te}_3$ and $\text{PbBi}_2\text{Te}_4$ \cite{hsieh_tunable_2009,kuroda_experimental_2012-1}. The TSS is robust against scattering from non-magnetic perturbations. Moreover, it possesses a helical character, which infers a defined spin polarization for a particular momentum value. These characteristics might lend themselves to a variety of new applications, especially, in spintronics, where transport and manipulation of spin currents at high temperatures and with diminutive scattering interactions are sought out \cite{hasan_colloquium:_2010,kane_topological_2011,qu_aharonov-casher_2011,zcaronuticacute_spintronics:_2004}.

It has already been shown that the modification of the $\text{Bi}_2\text{Se}_3$ surface with non-magnetic adsorbates
\cite{hsieh_tunable_2009,chen_massive_2010,valla_photoemission_2012,bianchi_simultaneous_2011,Benia_reactive_2011}, as well as heating up to 400 K \cite{hatch_stability_2011} does not alter TSS protection. However, it dopes the TSS and can even change surface electronic properties. Surface $n$-doping creates two sets of new states at the surface, which appear simultaneously in the immediate vicinity of the TSS within the projected bulk conduction band (CB) and valence band (VB) regions. While parabolic bands with a large Rashba splitting are observed above the Dirac point ($E_D$), the bands below are M-shaped, can overlap with the TSS, and do not show a clear spin splitting. These two band sets have been mainly interpreted as quantized subbands resulting from the confinement of a pair of two-dimensional electron gases (2DEGs) at the surface created via CB and VB bending \cite{bianchi_simultaneous_2011,Benia_reactive_2011,king_large_2011,bianchi_coexistence_2010,bahramy_emergent_2012}. Practically, the largely spin-split bands add a new feature to $\text{Bi}_2\text{Se}_3$ surface for spintronic applications \cite{king_large_2011}. Yet, the reason behind the discrepancy in Rashba splitting between CB and VB subbands is still not clear. An analogy with the case of 2DEG formation at InAs(111) and CdO(001) surfaces would suggest the splitting discrepancy to be due to a stronger band bending magnitude at CB than at VB \cite{king_surface_2010}. Here, by analyzing recorded surface band structures using angular resolved photoemission spectroscopy (ARPES) from differently treated $\text{Bi}_2\text{Se}_3$ surfaces, we examine the formation of the M-bands and their overlap with the TSS via band bending. We show that the potential gradients at both 2DEGs are similar and therefore not responsible for the splitting discrepancy. On the other hand, our first principles calculations show that the contribution of Bi 6$p$ states to VB is notably smaller than to CB. As the Bi 6$p$ states are characterized by a strong spin-orbit coupling (SOC), we hence attribute the spin-splitting discrepancy to a difference in SOC strength.

The $\text{Bi}_2\text{Se}_3$ crystal was grown following a vertical Bridgman method (see details in \cite{Epaps}). The ARPES measurements were done with a hemispherical SPECS HSA3500 electron analyzer characterized by an energy resolution of about 10\,meV. Monochromatized HeI (21.2\,eV) radiation was used as a photon source. During the measurements the vacuum pressure was less than $3\times 10^{-10}$\,mbar. The crystal was cleaved in vacuum at $2\times 10^{-7}$\,mbar or in air. When cleaving in air, the crystal was immediately put into a load-lock.

Fig.\ref{fig:BiSeWaterExp} shows the second-derivative of experimental surface band structures of $\text{Bi}_2\text{Se}_3$ obtained at different surface conditions. In Figs.\ref{fig:BiSeWaterExp}a and \ref{fig:BiSeWaterExp}b the crystal was cleaved in air then held in UHV for two and eight hours at 100 K, respectively. In Figs.\ref{fig:BiSeWaterExp}c and \ref{fig:BiSeWaterExp}d the crystal was cleaved in vacuum then exposed to 720 L and 1140 L of water vapor at 100 K, respectively (L : Langmuir)\cite{Epaps}. In all band structures, three main conduction quantized subbands (CSB) as well as three quantized valence subbands (VSB) induced via a surface reaction with water vapor are observed in addition to the TSS \cite{Benia_reactive_2011}. Depending on the surface reaction with the adsorbates, the bending of the conduction band minimum (CBM) varies and thus the CSBs get different Rashba splitting and energy positions \cite{bianchi_simultaneous_2011,Benia_reactive_2011}. The Rashba splitting is confirmed by the concentric contours in the energy cuts at 30 meV shown in Fig.\ref{fig:BiSeWaterExp}e. The VSB energies are also found to depend on time at low temperature and water exposure, but their M-shaped form remains unaffected. The M-like dispersion shows a band anisotropy, which is visible in the constant energy cuts at 645 meV and 700 meV, where the VSB contours are hexagonally warped (Fig.\ref{fig:BiSeWaterExp}e). The outer contour of C1 is also hexagonally warped near the Fermi level \cite{kuroda_hexagonally_2010,fu_hexagonal_2009}, however, here the warping is rotated by $30^{\circ}$ compared to VSB contours. The warping of the TSS follows C1 warping above $E_D$. Below $E_D$ the spectral intensity of TSS becomes very weak as the TSS vanishes rapidly in VB. Nevertheless, based on the low-energy model Hamiltonian of the TSS \cite{liu_model_2010}, we find the warping of the Dirac cone below $E_D$ is also rotated by $30^{\circ}$ compared to the warping above $E_D$, see Fig.\ref{fig:BiSeWaterExp}f. This difference in warping affects the out-of-plane spin component of both the quantized subbands and the TSS,
and might have important implications for the occurrence of symmetry breaking states on the surface of Bi$_2$Se$_3$  \cite{fu_hexagonal_2009,hasan_warping_2009,bahramy_emergent_2012}.

\begin{figure}
\centerline{ \includegraphics[width = 0.999\columnwidth]{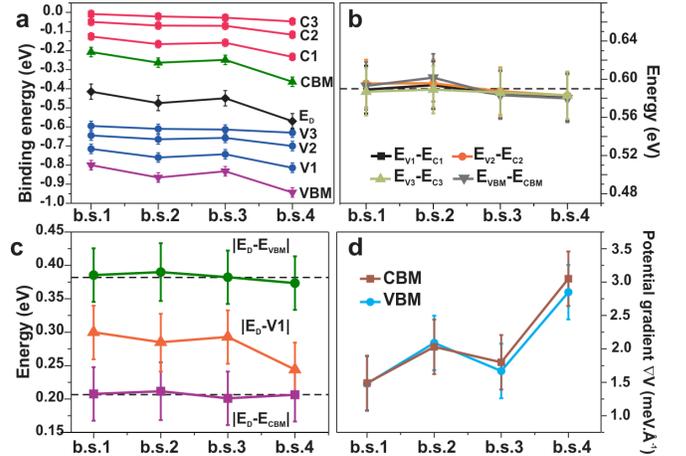}}
\caption{\textbf{a}. Fitted energy positions of CSBs, VSBs and $E_D$ and the extracted CBM and VBM energies at $k_\|=0$ $\text{\AA}^{-1}$ from surface band structures shown in Fig.\ref{fig:BiSeWaterExp}a, b, c, and d indexed here with b.s.1, b.s.2, b.s.3, and b.s.4, respectively. \textbf{b}. The difference in energy positions between paired CSB and VSB as well as between CBM and VBM at the surface. \textbf{c}. Variation of CBM, VBM and V1 with respect to $E_D$.  \textbf{d}. Variation of surface potential gradients generated by CBM and VBM bending. The triangular-well model was use to extract VBM, CBM, and  $\nabla V$ data (see text).} \label{fig:FitResults}
\end{figure}

The energy positions of $E_D$, CSBs, and VSBs in Figs.\ref{fig:BiSeWaterExp}a to \ref{fig:BiSeWaterExp}d are summarized in Fig.\ref{fig:FitResults}. The different subbands as well as $E_D$ follow a similar general trend. The VSB positions are closely associated to the CSB positions, whereby the subband pairs V1 and C1, V2 and C2, and V3 and C3 are evolving in parallel. As shown in Fig.\ref{fig:FitResults}b, all three subband pairs are separated by the same energy difference, which stays constant over the changes of the band structure induced by surface perturbation. An identical constant energy separation between the subband pairs has been extracted from all our measured ARPES data (not all shown here) as well as from all $\text{Bi}_2\text{Se}_3$ band structures found in the literature where the VSBs are shown \cite{valla_photoemission_2012,bianchi_simultaneous_2011,bahramy_emergent_2012,wray_topological_2011,bianchi_robust_2012}. This energy offset is equal to 590 meV $\pm$ 10 meV.
\begin{figure}
\centerline{ \includegraphics[width = 1\columnwidth]{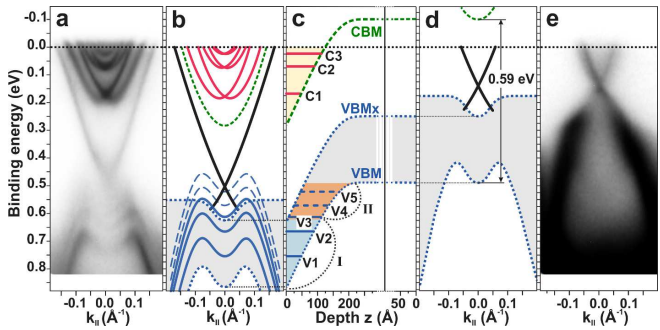}}
\caption{Sketch of surface bending of CB and VB and the formation of quantized CSBs and VSBs. \textbf{a}. Experimental band structure (Fig.1.c). \textbf{b}. Schematic outline of a. \textbf{c}. Representation of the surface with (left) and without (right) surface band bending at $k_\| = 0$ $\text{\AA}^{-1}$. \textbf{d}. schematic outline of e. \textbf{e}. experimental band structure directly after crystal cleaving in vacuum.} \label{fig:SketchBandBending}
\end{figure}
The parallel evolution of each subband pair strongly supports the fact that they are both a consequence of a similar band bending. Though the occurrence of the CSBs can easily be understood with a picture of a 2DEG confined between the CBM and the surface \cite{bianchi_simultaneous_2011,Benia_reactive_2011,bahramy_emergent_2012,king_large_2011}, the  simultaneous creation of VSBs via band bending is not straightforward. It could only be possible in a bending configuration comparable to the CBM. Consequently, it is necessary to have an energy gap within the surface projected bulk valence band, the minimum of which (VBM) should not be far below V1. Such a gap is predicted by first-principles calculations \cite{xia_observation_2009,zhang_first-principles_2010,eremeev_effect_2010}. In order to extract the CBM position and the surface potential gradient for the different band structures of Fig.\ref{fig:BiSeWaterExp}, a triangular potential well was used to model the CBM bending \cite{Benia_reactive_2011,ando_electronic_1982,davies_physics_1998}. The CSB positions at  $k_\|=0$ $\text{\AA}^{-1}$ obtained after fit of CSB with parabolic dispersive curves were taken as input parameters \cite{Benia_reactive_2011}. Similarly, VBM positions and surface potential gradients were extracted and are presented together with those for CSB in Fig.\ref{fig:FitResults}. As the bulk band quantization via bend bending occurs perpendicular to the surface, the effective mass $m^{*}$ of the band dispersion along $k_z$ ($\Gamma Z$ direction in the 3D BZ) is used in the triangular-well model. The fit with parabolic curves of the experimental CB and VB dispersion along $k_z$ from ref.\cite{bianchi_coexistence_2010} gives a value of 0.30$m_e$ $\pm$0.05$m_e$ for both bands. $m^{*}$ in parallel momentum is found to be smaller but similar for both subband sets with a value of 0.20$m_e$ $\pm$0.01$m_e$ at the CSBs $m^{\ast}$ and 0.19$m_e$ $\pm$0.03$m_e$ around $k_\|=0$ $\text{\AA}^{-1}$ (before $m^{\ast}$ changes sign) at the VSBs.
As shown in Fig.\ref{fig:FitResults}, the VBM follows the variation of the CBM, whereby $E_\text{CBM}-E_\text{VBM}$ is matching the subband pairs' energy separation in all cases. The two potential gradients are nearly identical and the energy separation between CBM and $E_D$, and, VBM and $E_D$ remains almost unchanged revealing an unaffected bandgap size as a function of band bending, in contrast to InAs and CdO cases \cite{king_surface_2010}.
\begin{figure*}
\centerline{\includegraphics[width = 1\textwidth]{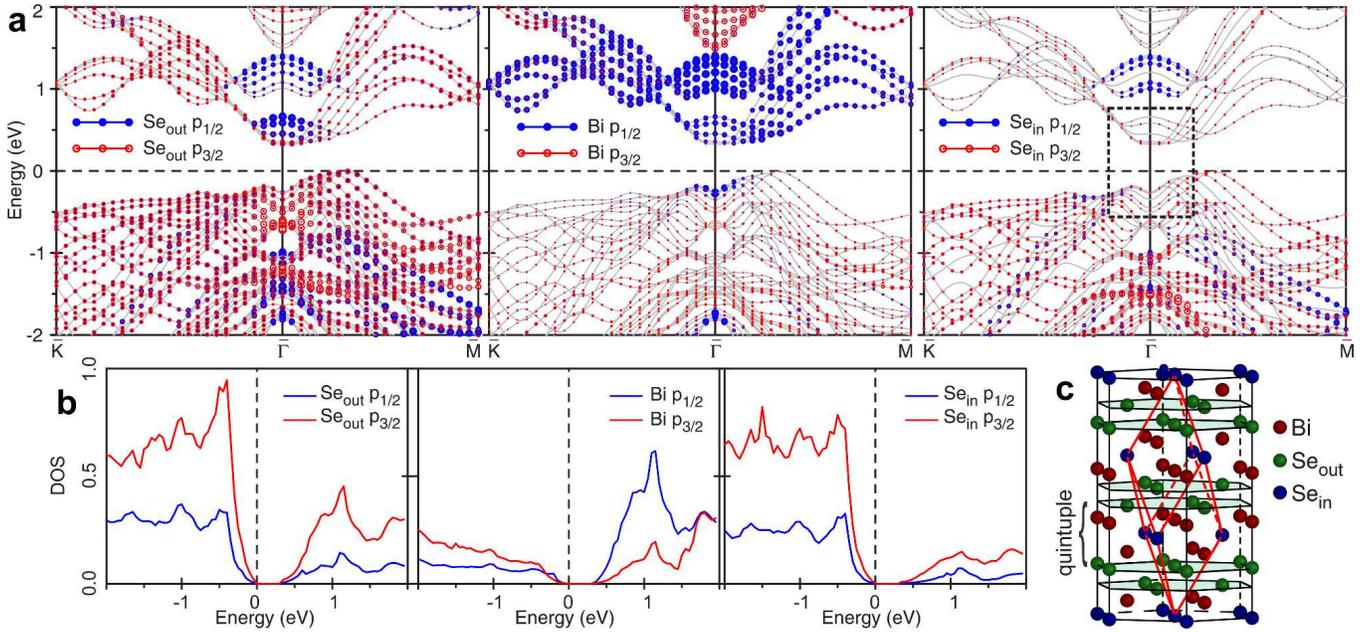}}
\caption{\textbf{a.} Fat bands representation of the projected band structure of $\text{Bi}_2\text{Se}_3$ along high-symmetry lines of a 2D BZ. The size of filled and open circles is proportional to the weight of $p_{1/2}$ and $p_{3/2}$ states of $\text{Se}_\text{out}$ (left), Bi (middle), and $\text{Se}_\text{in}$ (right) in the corresponding Bloch wave function. The projection onto a 2D BZ was simulated by plotting together bands with 6 equidistant $k_z$ values.} \label{fig:theory}
\end{figure*}
Accordingly, and as sketched in Fig.\ref{fig:SketchBandBending}, a downward band bending of VB parallel to the CBM creates a quantum well (QW). Yet, the presence of the valence band maximum (VBMx) as an additional barrier makes the created QW have special boundary conditions. The situation is more perceptible for a relatively strong band bending (see Fig.\ref{fig:SketchBandBending}). The confining well at the VB side could be viewed as a quantum well with two different regions: region I delimited by the crystal surface on one side and the VBM on the other side; and region II delimited by the VBMx on one side and the VBM on the other side. The resulting VSBs from region I are bound to the surface, and therefore, they should always appear below the TSS in the measured band structure whatever the band bending is, as it causes a rigid shift of TSS, CBM, VBMx and VBM. VSBs from region II are formed deeper in the bulk, and hence, they can overlap in the spectra with the TSS in the projected band structure under strong band bending. While only VSB below the TSS are observed here, a situation where one VSB overlaps with the TSS has been clearly observed in Ref. \cite{bianchi_simultaneous_2011} when excitation photon energy of 16 eV is used. The overlap of VSBs with the TSS constitutes an alternative explanation to the apparent time-reversal symmetry breaking and bandgap opening at $E_D$ in the case of Fe on $\text{Bi}_2\text{Se}_3$ \cite{bianchi_simultaneous_2011,wray_topological_2011, honolka_-plane_2012}.
On the other hand, the lowest M-shaped band (V1) in Ref. \cite{bianchi_simultaneous_2011} is regarded as being a surface state rather than a quantized VSB, as it falls below an estimated position of VBM. Assuming V1 as a surface state suggests that it evolves in parallel with $E_D$ position for the different band bending magnitudes. However, this is not the case, since the energy separation $|E_D-\text{V1}|$ does not remain constant, as shown in Fig.\ref{fig:FitResults}c.

According to the Rashba-Bychkov model \cite{bychkov_properties_1984} the spin-degeneracy can be lifted for free electron like states in a confined 2DEG \cite{bychkov_properties_1984,winkler_spin-orbit_2003}. This effect, which has been initially observed at 2D systems in semiconductor heterojunctions, plays an important role in the field of spintronics \cite{zcaronuticacute_spintronics:_2004,bychkov_properties_1984,winkler_spin-orbit_2003, datta_electronic_1990,wunderlich_experimental_2005,wu_spin_2010}. The Rashba-Bychkov model attributes the splitting effect to the combination of (1) breaking the inversion symmetry by an asymmetric confining potential at the surface or at the interface, and (2) SOC effects, which are inherent to the host semiconductor and/or induced by an effective potential gradient \cite{bychkov_properties_1984,wu_spin_2010}. Rashba splitting is explicit here at the CSBs. The VSBs, which are characterized by an $m^{\ast}$ similar to CSBs (see above), do not show any clear splitting, although both 2DEGs are under the same potential gradient, as shown above (Fig.\ref{fig:FitResults}). This splitting discrepancy observed simultaneously at the same surface is principally a direct proof that the potential gradient is a necessary-but-not-sufficient condition to spin-split the 2DEG bands. This suggests that different atomic SOC strength at both subband sets is responsible for the splitting difference \cite{petersen_simple_2000,ast_sp-band_2012}. The $p$ valence orbitals of Se(4$p$) and Bi(6$p$) atoms are characterized by dramatically different SOC. While for Se(4$p$) SOC parameter is 0.22 eV, for Bi(6$p$) it is five times larger with a value of 1.25 eV \cite{wittel_atomic_1974}. In the crystal, the predominant bonding character is a polar covalent $pp\sigma$ type between Bi and Se with charge transfer from Bi to Se \cite{mishra_electronic_1997}. At first glance, one could adopt a basic picture of energy bands, where the filled states (VB) and the empty states (CB) near the band gap are of Se(4$p$) and Bi(6$p$) character, respectively, and argue the splitting discrepancy with the SOC difference. However, such a description is oversimplified. It neglects the band inversion that characterizes the topology of $\text{Bi}_2\text{Se}_3$ at the $\Gamma$ point, which restructures CB and VB in the vicinity of the band gap \cite{zhang_topological_2009-1,zhang_first-principles_2010} and could therefore result in strong Bi contribution to VB.
In order to check the different contributions of the Bi and Se $p$ orbitals, we have performed first principles calculations of $\text{Bi}_2\text{Se}_3$ band structure using the fully relativistic linear muffin-tin orbital (LMTO) method \cite{andersen_linear_1975,perlov_py-lmto}. This implementation of the LMTO method uses four-component basis functions constructed by solving the Dirac equation inside an atomic sphere \cite{nemoshkalenko_relativistic_1983}, which is crucial for a correct description of $p_{1/2}$ states of heavy elements such as Pb or Bi \cite{macdonald_linearised_1980}. The self-consistent calculations were performed for bulk $\text{Bi}_2\text{Se}_3$ with rhombohedral unit cell using experimental lattice constants \cite{nakajima_crystal_1963}. The results of the calculations are summarized in Fig.\ref{fig:theory}. To compare with the experimental data, we will focus on the
energy/momentum region delimited by the dashed rectangle in the right panel. The bands near the top of VB are formed mainly by the $p_{1/2}$ and $p_{3/2}$ states of the outer (Se$_\text{out}$) and inner (Se$_\text{in}$) Se atoms of a
quintuple layer. Because of the band inversion, Bi $p_{1/2}$ states also contribute to the valence bands. This contribution is, however, restricted to the topmost bands (close to the bulk $\Gamma$-point \cite{Epaps}) in a small region of $k_\parallel$ around the $\bar{\Gamma}$ point. The bands near the bottom of CB are dominated by Bi $p_{1/2}$ states hybridized with the $p$ states of Se$_\text{out}$. Se$_\text{in}$ $p$-states contribute to the bands at about 1\,eV, whereas Bi $p_{3/2}$-states form bands above $\sim$1.5\,eV. The dominance of the Bi $p$-states in the CB bands suggests that the states at CB are subject to strong SOC. In contrast, the states in VB, especially near the VBM (close to the bulk Z-point \cite{Epaps}), experience weak SOC as they are dominated by Se $p$-states. Hence, the Rashba splitting in the CSBs is expected to be larger than in the VSBs, which explains the discrepancy observed in the experimental band structures. In addition, the M-shape of the VSBs, which is imposed by the VB dispersion, is expected to further decrease the Rashba splitting. Nonparabolicity effects have been found to reduce considerably the Rashba splitting especially for semiconductors with small bandgap, as it is the case for $\text{Bi}_2\text{Se}_3$ \cite{hu_zero-field_1999,matsuyama_rashba_2000,isic_nonparabolicity_2009}.

In summary, we examined the formation of two sets of quantized subbands at $\text{Bi}_2\text{Se}_3$ surface and discussed their discrepancy in Rashba splitting observed on experimental surface band structures. All the subbands are treated as resulting from bending of the conduction and the valence bands at the surface. The overlap situation of valence subbands with the topological surface state under strong band bending is discussed. Moreover, the band bending magnitude is found to be the same at both band sides, which makes the two sets of subbands evolve in parallel. LMTO band structure calculations reveal weak contribution of Bi 6$p$ states that are characterized by strong spin-orbit coupling to the valence band in comparison to the conduction band. Therefore, the discrepancy in the Rashba splitting is not due to a difference in potential gradients, but rather, to different spin-orbit coupling strength at both band sites.

A.\ Y.\ acknowledges the hospitality of Donostia International Physics Center (DIPC) in Donsotia/San Sebastian during his stay there. C.\ R.\ A.\ acknowledges funding from the Emmy-Noether-Program of
the Deutsche Forschungsgemeinschaft (DFG).

\end{document}